\def\year{2021}\relax
\documentclass[letterpaper]{article} % DO NOT CHANGE THIS
\usepackage{aaai21}  % DO NOT CHANGE THIS
\usepackage{times}  % DO NOT CHANGE THIS
\usepackage{helvet} % DO NOT CHANGE THIS
\usepackage{courier}  % DO NOT CHANGE THIS
\usepackage[hyphens]{url}  % DO NOT CHANGE THIS
\usepackage{graphicx} % DO NOT CHANGE THIS
\usepackage{natbib}  % DO NOT CHANGE THIS AND DO NOT ADD ANY OPTIONS TO IT
\usepackage{caption} % DO NOT CHANGE THIS AND DO NOT ADD ANY OPTIONS TO IT
\usepackage{stmaryrd,amsbsy,enumerate}
\usepackage{mdwlist}
\usepackage{pdfsync}
\usepackage{marginnote}
\usepackage{xspace}
\usepackage{latexsym,graphicx,amssymb,proof}
\usepackage{tabularx}
\usepackage{amsmath}
\usepackage{times}
\usepackage{color}
\usepackage{algorithm}

\usepackage{tikz}
\usetikzlibrary{arrows}

\newtheorem{definition}{\vspace{1mm}Definition}[section]

\newtheorem{example}[definition]{\vspace{1mm}Example}

\newtheorem{theorem}[definition]{\vspace{1mm}Theorem}

\newtheorem{proposition}[definition]{\vspace{1mm}Proposition}

\newenvironment{proof}{\smallskip\noindent\textit{Proof:}}{\hfill$\square$\smallskip}

\newcommand{\red}[1]{\textcolor{red}{#1}}

\newcommand{\violet}[1]{\textcolor{violet}{#1}}
\newcommand{\brown}[1]{\textcolor{brown}{#1}}

\newcommand{\ignore}[1]{}
\newcommand{\WJ}[1]{{\color{blue} #1}}
\newcommand{\GB}[1]{\violet{#1}}
\newcommand{\para}[1]{\smallskip\noindent\textbf{#1}}

 \newcommand{\nats}{\mathbb{N} }
 \newcommand{\reals}{\mathbb{R} }
  \newcommand{\set}{\mathcal{X} }
  \newcommand{\rvx}[1]{X^{(#1)} }  %random variable X with upper index
   \newcommand{\prob}{\textit{Pr}}
  \newcommand{\tm}{{\bf P}}  %transition matrix P
  \newcommand{\tme}{P}  %transition matrix entry P with indeces ij
  \newcommand{\pv}[1]{{\bf p}^{(#1)}}  %probability vector p with upper t time index
    \newcommand{\pve}[1]{p^{(#1)}}  %probability vector entry p with upper t and lower  i index
    \newcommand{\sd}{\bar{\bf p}}  %stationary distribution
        \newcommand{\sde}{\bar{p}}  %stationary distribution

\newcommand{\cans}{\mathcal{V\!\!\;O}}    % votable options (candidates)
\newcommand{\vots}{\mathcal{V}}       % voters
\newcommand{\pref}{\prec}             % preference relation
\newcommand{\ord}{\mathcal{P}\!\mathit{ref}}     % set of preference relations (orderings)
\newcommand{\out}{\mathcal{O}\!\mathit{ut}}      % election outcomes
\newcommand{\vex}{\mathcal{X}}      % election outcomes
\newcommand{\scf}{{C}}        % social choice function

\newenvironment{mytikzgraph}
  {\begin{tikzpicture}[->,>=stealth',shorten >=1pt,auto,node distance=2.2cm,thick,main node/.style={circle,draw,font=\sffamily\bfseries}]}
  {\end{tikzpicture}}
\tikzstyle{belowleft}=[below left of=a,below=10pt]
\tikzstyle{belowright}=[below right of=a,below=10pt]
\tikzstyle{bendleft}=[bend left]

\title{Convergence Voting: From Pairwise Comparisons to Consensus}
\author{Gergei Bana,$^{\rm 1}$ Wojciech Jamroga,$^{\rm 2,3}$ David Naccache,$^{\rm 4}$ and Peter Y. A. Ryan$^{\rm 2}$}
\affiliations{
    \smallskip
    \textsuperscript{\rm 1} University of Missouri at Columbia, Columbia MO, USA, \texttt{\small bana@math.upenn.edu}\\
    \textsuperscript{\rm 2} SnT, University of Luxembourg, Esch-sur-Alzette, Luxembourg, \texttt{\small\{wojciech.jamroga,peter.ryan\}@uni.lu}\\
    \textsuperscript{\rm 3} Institute of Computer Science, Polish Academy of Sciences, Warsaw, Poland\\
    \textsuperscript{\rm 4}  ÉNS (DI), Information Security Group, CNRS, PSL Research University, 75005, Paris, France\\ \texttt{\small david.naccache@ens.fr}
}

\begin{document}

\maketitle
\begin{abstract}
An important aspect of AI design and ethics is to create  systems that reflect aggregate preferences of the society. To this end, the techniques of social choice theory are often utilized.
We propose a new social choice function motivated by the PageRank algorithm. The function ranks voting options based on the Condorcet graph of pairwise comparisons. To this end, we transform the Condorcet graph into a Markov chain whose stationary distribution provides the scores of the options.
%, and the candidates are ranked from the highest to the lower scores.
We show how the values in the stationary distribution can be interpreted as quantified aggregate support for the voting options, to which the community of voters converges through an imaginary sequence of negotiating steps. Because of that, we suggest the name ``convergence voting'' for the new voting scheme, and ``negotiated community support'' for the resulting stationary allocation of scores.

Our social choice function can be viewed as a consensus voting method, sitting somewhere between Copeland and Borda. On the one hand, it does not necessarily choose the Condorcet winner, as strong support from a part of the society can outweigh mediocre uniform support. On the other hand, the influence of unpopular candidates on the outcome is smaller than in the primary technique of consensus voting, i.e., the Borda count. We achieve that without having to introduce an ad hoc weighting that some other methods do.
\end{abstract}

%\keywords{social choice, consensus voting, Markov chain, iterative decision making}

%\maketitle

%%%%%%%%%%%%%%%%%%%%%%%%%%%%%%%%%%%%%%%%%%%%%%%%%%%%%%%%%%%%%%%%%%%%%%%%%%%%%%%%%%%%%%%%%%%%%%%%%%%%%%%%%
\section{Introduction}\label{sec:intro}

%\red{Target conference: Conference on Artificial Intelligence, Ethics and Society deadline Jan 31, 2021 - Page limit: 6-10 pages + references}
%\red{Introduce rank centrality earlier, and then the construction steps. Remove the disappointment factor.}
%\red{We do not entirely throw away the information of individual ranking, it is just thrown away at at an earlier stage. Borda also throws away once you start adding.}

%\red{Section 8 in Handbook of comsoc. Reason with consensus classes and distance metric.}

Voting is important to the human society, as many collective decisions are made by means of elections and referenda~\cite{Brandt16comsoc-handbook,Hao17evoting}.
Voting-based mechanisms are also used in design of artificial intelligence  systems to aggregate goals of individual agents into a cohesive collective decision~\cite{Weiss99mas},
or to design AI systems that work in accordance with aggregate preferences of the society~\cite{Baum20}.
There is a multitude of different aggregation schemes, called social choice rules, each answering different needs~\cite{Arrow02socialchoice-handbook,Shoham09MAS,Brandt16comsoc-handbook}.

\para{Voting and Consensus.}
It has been known for almost 70 years that no social choice rule can satisfy all the desirable theoretical properties of non-dictatorship, universality, independence of irrelevant alternatives,
%monotonicity, non-imposition,
and unanimity~\cite{Arrow50impossibility}.
Thus, the designer of a voting scheme must decide which properties are more dispensable.
Perhaps more importantly, practical political and social concerns lead to different solutions, from rules that favor the broadest possible representation of the society and protection of minorities, to ones that focus on the effectiveness of the elected body and usually favor the majority. The problem at hand may be even subtler when a candidate's goal is \emph{not} to win the election but rather to lose by a slight margin.\footnote{E.g. a region can have no economical interest to gain independence, but wishes to demonstrate to the central government that independence is feasible so as to obtain wider autonomy. }

The idea of \emph{consensus voting rules} is to somehow reconcile the two extremes. Typical examples of consensus-oriented rules include Condorcet voting and variants of Borda, but one can argue that Single Transferable Vote (STV) and Instant-Runoff Voting (IRV) are driven by similar concerns~\cite{Robert11rulesOfOrder}. Even the two-round system used in many presidential elections (also known as the second ballot) can be seen as a crude attempt to balance the will of majority with the breadth of influence.

\para{Convergence Voting.}
Our starting point is a collection (complete or not) of pairwise comparisons of voting options by the voters. Pairwise comparisons are not only important in voting schemes but are also very suitable for a machine learning setting \cite{FurHul20}.
This paper is based on the observation that, in human communities, consensus is often reached iteratively rather than in a single step, through negotiations that eventually converge to some agreement \cite{heg:kra:02}.
%People incrementally shift their opinions and preferred choices in a multi-step process that (hopefully) converges to a solution that (more or less) satisfies everybody.
Following that intuition, we propose a  procedure that takes as input the matrix (or, equivalently, graph) of pairwise comparisons between voting options, and produces a distribution of weights that we call scores, which is reached via an imaginary sequence of negotiating steps in the community of voters. Each score represents the aggregate popularity of the associated option.
%In the standard case, the input matrix is just the Condorcet graph, but in general it can be any matrix/graph that shows how much one voting option is favored over another by the population of voters.
Our main contributions are: the definition of the voting procedure, its interpretation as a \emph{negotiated community support}, and the description of two infinite negotiating procedures both producing the same scores in their limits.

To this end, we propose a simple but non-trivial transformation of the graph into a discrete Markov chain.
That is, transition probabilities between the voting options are assigned to the graph based on the number of voters who prefer one option over another.
The uniform probability distribution over the options in the graph can be seen as representing a nondiscriminatory initial ranking of the options. The chain then determines a unique sequence of distributions as the transition matrix is applied on the initial distribution once, twice and so on, corresponding to the steps of the negotiation process. For
%a
our Markov chain, this always converges to a \emph{stationary distribution}, which defines the output of our procedure. Namely, the score of each option is defined as its probability in the limiting  stationary distribution.
%of  the Markov chain obtained from the initial uniform distribution (though under certain conditions the initial distribution does not matter).

The scores can be used in various ways, such as the voting option with the highest score can be chosen as the winner, or, for elections of collective bodies (e.g., the parliament), the scores could  define the shares of different parties in the composition of the body, and so forth.

% which can be used to determine the winning option in at least three ways.
%First, the distribution can serve as a ranking of voting options, with the highest scored option selected as the winner.
%Secondly, the winner can be drawn at random from among the options according to the output distribution.
%Thirdly, for elections of collective bodies (e.g., the parliament), the distribution can be used to define the shares of different parties in the composition of the body. DN: this is very interesting. I really think that this should be mentionned

\para{Convergence Voting as Imaginary Negotiation Process.}
The new social choice function\footnote{Throughout the paper, we will use the terms \emph{social choice rule}, \emph{voting rule}, and \emph{social choice function} interchangeably.}
is supposed to capture the spirit of iterative mutual adjustments in the society.
We provide two related views of this imaginary negotiation process: one based on iterative reallocation of community support, and the other one in terms of iterated probabilistic change of the selected decision.
According to the former, the negotiations transform a quantified \emph{negotiated community support} for the voting options. The process starts with an even distribution of support. Then, at each negotiation round, every voter has an equal share of the support value on each option, which she can further divide into equal parts, and redistribute over the other options. This redistribution
of the shares results in a new assignment of
%shares
support, from which the next round can be executed in the same manner.
%Continuing this ad infinitum, this process at each step reproduces the  distributions of the Markov chain at each step, and hence it also converges to the same distribution.
As the limiting stationary distribution is stable under the above rearrangement of shares,
%if the above redistribution of the preferences is applied on it, we get  back the same distribution, and hence
it can be considered as a consensus.

In the probabilistic interpretation, we assume that, at each round, a single option is contemplated by the society as a tentative winner.
The very first option is chosen uniformly at random.
Then, a
%a random voter and another voting option are chosen (again uniformly)
voter and another voting option are chosen  uniformly randomly.
If the voter prefers the other option, the society moves on to contemplate the latter one, otherwise they stick with the previous one; this sequence of steps is repeated ad infinitum.
We notice that a random walk in the Markov chain corresponds to a sequence of shifts between one contemplated winner to another.
Thus, our score can be interpreted as the
%expected frequency with which the given option will occur in the sequence.
frequency with which the given option will occur in the sequence. By the ergodic theorem,
that frequency agrees with our score by probability $1$.
%\red{GB remark: it is not just expected frequency, it is much stronger, by probability $1$, the frequency converges to it. }
%More precisely, in each round, there is a "preferred option". At each round, the options then can be ranked by the frequency they have been preferred options.
%In other words, at each round, a randomly chosen voter can determine whether to override the preferred option or not by a randomly chosen other option, and the frequency by which each option has been chosen as preferred option can be recorded.
%Then by the ergodic theorem, by probability $1$, the frequency of any option's  occurrence as preferred option in an infinite number of iterations  converges to  our score of that option.

\para{Computing the Winner(s).}
Convergence voting has two welcome computational properties.
First, convergence is guaranteed for any pairwise comparison graph given as input.
Secondly, the output distribution is very easy to compute. In fact, there is no need to carry out the actual iterations, as the distribution can be obtained by solving a simple set of linear equations. %as well known from the theory of Markov chains.
Thus, one can determine the output without simulating the asymptotic convergence of opinions by which the rule is defined.

\para{Structure of the Paper.}
The paper is structured as follows. In Section \ref{sec:preliminaries} we summarize the background we need from
the theory of social choice as well as the basics of the theory of Markov chains. In Section \ref{sec:convvoting}, we motivate and define our new voting scheme. In Section \ref{sec:interpretations} we show how convergence voting can be considered as a consensus reached by an infinite number of negotiating steps. Here we also introduce our notion of a negotiated community support. Finally in \ref{sec:further-properties}, we compare our scheme to other notable schemes and consider each of Arrow's properties. Besides the technical definitions, we have also provided a number of simple examples and verbal explanations to make it suitable to a multi-disciplinary audience.
%
%\red{TBD}

\subsection{Related Work}\label{sec:related}

We propose a social choice function based on a probabilistic model of possible iterated change of support for voting options.
In a way, this extends the idea behind the STV and IRV voting rules which also have iterative reassignment of votes, albeit in a very limited manner~\cite{Bartholdi91stv,Cary11irv,Robert11rulesOfOrder}.
More generally, iterative voting procedures have been explored e.g.~in~\cite{Meir10convergence,Lev12iterativeVoting,Grandi13iterative,Hassanzadeh2013consensus,Obraztsova15convergence}.
However, those approaches assume that the voting itself proceeds in rounds, and the participants can change their votes from one round to another.
This is typically modeled as a game in which the voters choose deterministic long-term strategies, and the convergence to a Nash equilibrium is studied.
In contrast, our approach assumes that the voters express their preferences once, and the outcome is defined by convergence of a \emph{virtual probabilistic procedure}.

In this sense, our proposal is closer to the work on iterative judgment aggregation in~\cite{Slavkovik16iterativeJA}, although the mathematical details differ completely (the procedure in~\cite{Slavkovik16iterativeJA} is not probabilistic in the first place).
Also, the work on viscous democracy~\cite{Boldi11viscous} comes close. The differences are as follows: (i) our model focuses on the flow of support between candidates, whereas~\cite{Boldi11viscous} is based on the flow of power between voters; (ii) we assume constant influence of the voter over the collective decision (i.e., single ballot), while they assume constant influence over peers (via the delegation factor); (iii) we model the flow of support  with a Markov chain, while~\cite{Boldi11viscous} uses arbitrary (non-normalized) weighted graphs to model the flow of influence.

Mathematically, our new voting rule has been inspired by the Google PageRank algorithm of ranking web pages~\cite{Brin98pagerank}. In particular, we follow the idea of transforming an arbitrary graph into a Markov chain, and using its stationary distribution to rank options.
Moreover, the mathematical structure of convergence voting is similar to rank aggregation algorithms for tournaments, based on statistical estimators, especially MC3~\cite{Dwork01mc3} and Rank Centrality~\cite{Negahban12rankcentrality,Negahban17rankcentrality}.
In fact, the output of our rule coincides with MP3/Rank Centrality on total preference profiles, i.e., when all voters have strict preference between each pair of candidates.
Note, however, that MP3/Rank Centrality have been proposed for a different purpose, namely to {statistically estimate an objectively existing} ``worth'' of goods on the market or players in a sport tournament. Consequently, we are driven by different intuitions, and we {construct our function based on} two arguments, all different from theirs. We are not looking for an estimator, but instead we are trying to obtain a good aggregation of preferences, independently of the voting options' objective qualities.
Even more importantly, our social choice function returns significantly different (and arguably more intuitive) output for partial preferences, i.e., when some voters are indifferent w.r.t. some voting options.

%our ranking turns out to be the same as the Rank Centrality estimator when the (objective) ranks are estimated from a statistic of pairwise wins.)

\section{Preliminaries}\label{sec:preliminaries}

\subsection{Voting Rules}\label{sec:socialchoice}

We  recall the notion of a social choice function that formalizes  collective decision making, usually realized by voting.

\begin{definition}[Social choice function]
	Let $\vots$ be the set of agents (also called \emph{voters}), $\cans$ the set of \emph{votable options},
	%(typically, candidates in an election)
	and $\ord$ a set of strict partial order relations over $\cans$. The relations are used to represent the voters' preferences over options.
	%A \emph{preference profile} $\PrefProf = \Pi_{v\in\vots} \ord\}$ specifies .
	Let $\out$ be the set of \emph{voting outcomes}.
	A \emph{social choice function} is a mapping $\scf : \ord^{|\vots|} \rightarrow \out$ that aggregates individual preferences of the voters into a collective decision.
	That is, for each \emph{preference profile} $(\pref_1,\dots,\pref_{|\vots|}) \in \ord^{|\vots|} $, the function returns an outcome $o\in\out$.
\end{definition}

Social choice functions provide an abstraction for the decision-making mechanism behind elections, referenda, plenary votes, etc. The voters express their preferences (usually, by filling in and casting their ballots), and the social choice function determines the outcome.
In case of an election, the voting options can be often identified with the set of candidates  (persons, parties, committees) that stand in the election. In case of referenda and plenary votes, the options are issues to be decided upon.
The voting outcomes are typically seen as either a preference relation over the voting options, a ranking of the options (e.g., the tally), or the option selected as the winner.
%In the latter case, it is often convenient to assume that $\cans = \out$.
One can also see this as a 2-stage process, with a ranking produced first from the ballots, and a winner (or winners) determined next according to the ranking.
This is the view we adopt in this paper.

Note that we allow for partial preferences. When a voter does not compare two options $A$ and $B$, we interpret that as the voter is \emph{indifferent} about which choice is better, $A$ or $B$.

%Example social choice functions include majority, plurality, Copeland method, and variants of Borda.

\begin{example}[Majority, plurality, and Borda]\em
	\label{ex:presidential}
	Consider a presidential election with 3 candidates: $\cans = \{A,B,C\}$ and 5 million voters, with the following distribution of preferences:\
	$B\pref C\pref A$ (1M voters),\ $C\pref A\pref B$ (1M voters),\ $A\pref B$ (1M),\ $B\pref C$ (1M),\ and $C\pref A$ (1M). Note that many voters express only partial preferences between the candidates.
	 $A$ and $B$ are most preferred by 2M voters each, and $C$ is most preferred by 1M voters.
	Thus the majority rule produces no winner (no-one is on the top of over 50\% of the {preference orderings}), and the plurality rule produces a tie between $A$ and $B$.
	The variant of Borda assigning the top candidate in a ballot with score $|\cans|-1$, the next one with $|\cans|-2$, etc. (unranked candidates getting $0$) yields the scores of 6M for $A$, 5M for $B$, and 4M for $C$, thus making $A$ win.
\end{example}

\subsection{Pairwise Comparison Graphs}\label{sec:pcgraphs}

Marquis de Condorcet postulated that if a candidate is preferred by the voters over any other candidate (in the sense that he/she would win a plurality election against every other candidate alone), the candidate should be the winner~\cite{Condorcet1785voting}.
%In a Condorcet voting, pairwise preferences are recorded. That is, how many voters prefer candidate A to candidate B, candidate B to candidate A, candidate A to candidate C, and so forth. Such record may come from preference lists of the voters, but not necessarily. There is a Condorcet winner, if there is a candidate who is preferred to each of the other candidates by more voter than the other is preferred to him/her.
The \emph{Condorcet winner} may not exist, and even if it exists it {is not always an obvious best choice}.
Nevertheless, Condorcet criterion draws attention to pairwise comparisons between voting options.
{They provide an important abstraction of a preference profile which helps to balance conflicting preferences of different voters and reach a consensus.
This can be conveniently represented by a graph.}

\begin{definition}[PC graph, Condorcet graph]
	A \emph{pairwise comparison graph (PC graph, for short)} is a weighted directed graph $G = (\vex,E)$ with no loops.
	The vertices correspond to the voting options, i.e., $\vex = \cans$. The edges are weighted by natural numbers, i.e., $E : \vex\times \vex \rightarrow \nats$ with $E(x,x) = 0$ for any $x\in \vex$.
	The weight $E(A,B)$ {on the edge from $A$ to $B$} represents the strength of preference for $B$ in comparison to $A$.
	%	An \emph{enhanced PC graph (EPC graph, for short)} is a pairwise comparison graph with added loops, i.e., it is not required anymore that $E(v,v) = 0$.
	
	Let $(\pref_1,\dots,\pref_{|\vots|}) \in \ord^{|\vots|} $ be a preference profile for the set of voting options $\cans$. The associated \emph{Condorcet graph} is a PC graph with vertices $\cans$ and
	$E(A,B)$ is the number of voters who prefer option $B$ over option $A$, that is the cardinality of the set $\{\ i\in (1,...,|\vots|) \ : \ A \pref_i B\ \}$.	
\end{definition}

\begin{figure}[t]
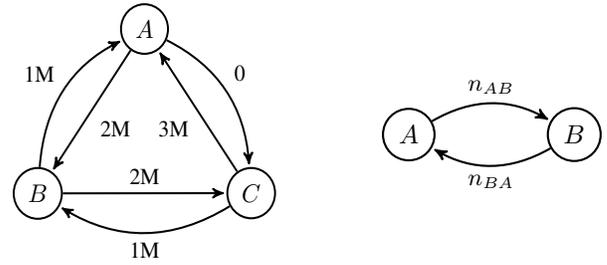
\centering
	\begin{tabular}{@{}c@{\qquad\qquad}c@{}}
		\begin{tabular}{@{}c@{}}
			\resizebox{3.5cm}{3.5cm}{%
			\begin{mytikzgraph}
				\input{graphABC}
			\end{mytikzgraph}}
		\end{tabular}
		&
		\begin{tabular}{@{}c@{}}
			\begin{mytikzgraph}
				\input{graphAB}
			\end{mytikzgraph}
		\end{tabular}
	\end{tabular}
	\caption{Condorcet graphs:\ (a) presidential election of Example~\ref{ex:presidential};\ (b) parliamentary election of Example~\ref{ex:parliamentary}}
	\label{fig:condorcet}
\end{figure}

\ignore{Other constructions are also possible -- for instance, the PC graph for cardinal voting (such as STAR voting~\cite{starvoting}) should label $(A,B)$ with the aggregate difference of scores for $B$ and $A$ for the voters who prefer $B$ over $A$.
Nevertheless, we focus on Condorcet graphs in this paper.}

\begin{example}\em
	\label{ex:condorcet-graph}
	The Condorcet graph for the preference profile of Example~\ref{ex:presidential} is shown in Figure~\ref{fig:condorcet}a.
	Clearly, the election has no Condorcet winner {($A$ is defeated by $B$, $B$ is defeated by $C$, and $C$ by $A$)}.
	On the other hand, $A$ seems intuitively the strongest option, as its win over $C$ is somewhat stronger than the wins of $C$ over $B$ and $B$ over $A$.
\end{example}

\begin{example}%[Another Condorcet graph]
\em
	\label{ex:parliamentary}
	Consider also another scenario, of a parliamentary election with only two parties, $A$ and $B$. Let $n_{AB}$ be the number of voters that prefer $B$ over $A$, and $n_{BA} > n_{AB}$ be the ones preferring $A$ over $B$.
		%The remaining $n_{BA} - n_{AB}$ voters are indifferent.
	The Condorcet graph for the election is presented in Figure~\ref{fig:condorcet}b.
	Here, $A$ is clearly the Condorcet winner. However, this does not answer the question \emph{how many} seats should be allocated to each party.
\end{example}

Inspired by the Google's PageRank algorithm~{\cite{Brin98pagerank}, and similarly to some previous works on rank aggregation~\cite{Dwork01mc3,Negahban12rankcentrality,Negahban17rankcentrality}, we will show how to transform the Condorcet graph into a Markov chain whose stationary distribution delivers a satisfactory  ranking of the voting options.}
%The way to do this however is not obvious.}

\subsection{Discrete-Time Markov Chains}\label{sec:markov}
%\subsection{Discrete-Time Time-Homogeneous Finite Markov Chains}\label{sec:markov}

We briefly summarize here what we need from the theory of Markov processes.
%	In this paper we need only special kinds of Markov Chains, which are categorized as discrete-time, time-homogeneous, with finite state space.
Let $\set=\{x_1,...,x_n\}$ be a finite state space. A \emph{discrete-time, time-homogeneous finite Markov chain} over $\set$ is a sequence of random variables $(\rvx t)_{t \in \nats}$ over some probability field $(\Omega,Pr)$, taking values in $\set$, such that for all $t\in \nats$,  $i,j,k_0,...,k_{t-1} \in \{1,...,n\}$,
\[\footnotesize{\begin{array}{l}\prob\Big(\rvx {t+1} = x_j \Big| \rvx t = x_i ,\rvx {t-1} = x_{k_{t-1}},..., \rvx {0} = x_{k_0} \Big)  \\ \hspace{1cm}=  \prob\Big(\rvx {1} = x_j \Big|  \rvx 0 = x_i\Big) = P_{ij}\end{array}}\]
%$\prob(\rvx {t+1}\! = s_j | \rvx t = s_i ,\rvx {t-1}\! = s_{k_{t-1}},..., \rvx {0} \! = s_{k_0} )  = \prob(\rvx {1} \! = s_j |  \rvx 0\! = s_i) = P_{ij}$
The process is called Markov because it is memoryless, as the above conditional probability does not depend on {the history} $x_{k_0}, ... , x_{k_{t-1}}$, only on {the current state} $x_i$ and {the target state} $x_j$. It is a chain because
$\set$ is discrete, it is finite because $\set$ is finite, it is discrete time because  $t$ is in $\nats$, and it is time-homogeneous because the
transition probability does not depend on $t$.

Such a process can be represented as a finite  directed labeled  graph with $\set$ as the set of vertices, $\set \times \set$ as the set
of directed edges $(x_i,x_j)$ pointing from $x_i$ to $x_j$ and labeled by $\tme_{ij}$ transition probability from $x_i$ to $x_j$. For all $t\in \nats$ and $i \in \{1,...,n\}$, let
 $\pve t_i:= \prob(\rvx {t} = x_i)$,  and let
$\pv t$ denote the row vector $(\pve t_1, ..., \pve t_n)$. Then with the transition matrix $\tm := (\tme_{ij})_{i,j \in \nats}$, we have that
	$\pv{t+1} = \pv{t} \tm$
	for all $t\in \nats$, where $\pv t \tm$ is the row  vector $\pv t$ multiplied from the right with the matrix $\tm$. Clearly, $\tm^{(u)}:= \tm^u$ (the matrix $\tm$ multiplied by itself $u$ times) for some $u \in \nats$ gives the transition probabilities from $x_i$ to $x_j$ after $u$ number of steps, and  $\pv{t+u} = \pv{t} \tm^{(u)}$.

Since all rows of $\tm$ add to $1$, the column vector with $1$ in all entries is a right eigenvector with eigenvalue $1$. The left eigenvectors with eigenvalue $1$, whose components add up to $1$, are called stationary distributions of the Markov chain as they are invariant under the application of the transition matrix $\tm$. As $1$ is always an eigenvalue, there is at least one stationary distribution. When the transition matrix is such that for all $i,j\in \{1,...,n\}$ it allows
a transition from $x_i$ to $x_j$, and also from $x_j$ to $x_i$, with non-zero probability (that is, when there are some $u,v\in \nats$ such that $\tme^{(u)}_{ij} \neq 0$ and $\tme^{(v)}_{ji} \neq 0$ then the Markov chain is called irreducible. An irreducible finite Markov chain always has a unique stationary distribution $\sd$.  If furthermore the irreducible finite Markov chain is also aperiodic (that is, for all $i\in \{1,...,n\}$, for large enough $u \in \nats$, $\tme_{ii}^{(u)} >0$), then  $\lim_{t\rightarrow \infty} \pv t = \sd$ no matter what the $\pv 0$ initial distribution is, with the limit being taken component-wise. Furthermore, the ergodic theorem of Markov processes states that for irreducible aperiodic Markov chains, the probability of the set of all those infinite chains for which the number of visits to each state $x_i$ until step $t$ divided by $t$ converges to $\sde_i$, equals 1.

By renumbering the states, the transition matrix of any finite Markov chain can be brought to the \emph{canonical form}
\begin{equation}\label{eq:genmat}
\footnotesize{\tm = \begin{pmatrix}
\tm_1 & {\bf 0} & \cdots & {\bf 0} & {\bf 0} \\
 {\bf 0} & \tm_2 &\cdots & {\bf 0} & {\bf 0} \\
 \vdots & \vdots & \ddots & \vdots & \vdots \\
{\bf 0}& {\bf 0} &\cdots  &  \tm_m & {\bf 0}\\
 {\bf S}_1 & {\bf S}_2 & \cdots & {\bf S}_m &   {\bf Q}
\end{pmatrix}}
\end{equation}
Where $\tm_k$, ${\bf 0}$'s,
${\bf S}_k$, and  ${\bf Q}$ are also matrices, the ${\bf 0}$'s have all $0$ entries,
${\bf Q}$ has no closed communication class (any chain will lead out of the states on which ${\bf Q}$ acts by probability $1$), and each $\tm_k$ is irreducible.
Taking $\sd_k$ (where $k$ is a natural number between $0$ and $m$) to be the stationary distribution of $\tm_ k$, any stationary distribution of $\tm$ is a convex combination of the stationary vectors of the form $({\bf 0}, \sd_k, {\bf 0} )$, and again ${\bf 0}$ denotes various sized zero row vectors. If each $\tm_k$ is aperiodic, $\lim_{t\rightarrow \infty} \pv t$ is still  equal to some stationary distribution. The limiting transition matrix, that is, the matrix $\tm^\infty$ such that for any $\pv 0$,  $\lim_{t\rightarrow \infty} \pv t = \tm^\infty \pv 0$ is also of canonical form if $\tm$ is, with all row vectors of $\tm^\infty_k$ being $ \sd_k$, and ${\bf Q}^\infty = {\bf 0}$ on the diagonal, while ${\bf S}^\infty_k = ({\bf I} - {\bf Q})^{-1} {\bf S}_k \tm^\infty_k$.

\begin{figure}[t]
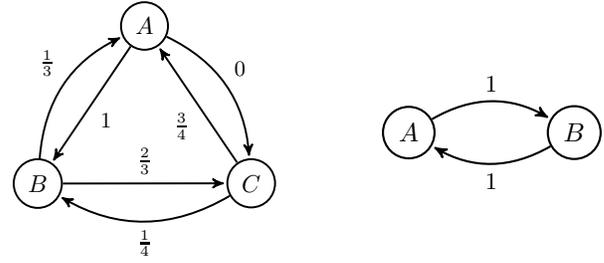
\centering
	\begin{tabular}{@{}c@{\qquad\qquad}c@{}}
		\begin{tabular}{@{}c@{}}
			\resizebox{3.5cm}{3.5cm}{%
			\begin{mytikzgraph}
				\input{markovchainABC}
			\end{mytikzgraph}}
		\end{tabular}
		&
		\begin{tabular}{@{}c@{}}
			\begin{mytikzgraph}
				\input{markovchainAB}
			\end{mytikzgraph}
		\end{tabular}
	\end{tabular}
	\caption{Normalized Condorcet graphs: \ (a) of Example~\ref{ex:presidential};\ (b) of Example~\ref{ex:parliamentary}}
	\label{fig:firsttry}
\end{figure}

\begin{example}[Markov chains for PC graphs]\em
	\label{ex:normalized}
	Given a pairwise comparison graph, one can transform it into a Markov chain by normalizing the weights of the outgoing edges for every vertex.
	The Markov chains obtained this way from the Condorcet graphs of Figure~\ref{fig:condorcet} are shown in Figure~\ref{fig:firsttry}.
\end{example}

\section{Convergence Voting}\label{sec:convvoting}

{Our social choice function is supposed to simulate an iterative process of reaching consensus among the voters. Thus, it makes sense} to use stable distributions on Condorcet graphs in order to rank the candidates.\footnote{
  Note that we focus on Condorcet graphs in the rest of the paper, but the method can be as well applied to PC graphs obtained in any other way.}
Clearly, the transition probabilities should be related to the weights in the PC graph: the stronger the preference for $A$ over $B$ is, the more likely the {shift} from $B$ to $A$ should be.
%We also assume that our model for the random walk is a Markov chain. As a consequence, the Markov stationary distribution gives us a ranking of the voting options.
If stronger preference means higher transition probability, than it is intuitive that the stationary distribution should rank the options according to how preferred they are.
However, it is not obvious how to exactly define the transition probabilities in the Markov chain.
In particular, the simplest solution of using normalized Condorcet graphs does not work.
%We rectify the procedure by enhancing the PC graphs with loops in Sections~\ref{sec:convvoting-second} and~\ref{sec:convvoting-final}.

%Having obtained a satisfactory construction that produces a ranking of voting options, we also present in Remark~\ref{sec:election-outcome} how they can be used to determine the winner(s) of the vote.

The mathematical structure of convergence voting turns out to be similar to the MP3/Rank Centrality aggregators~\cite{Dwork01mc3,Negahban12rankcentrality,Negahban17rankcentrality}. We will discuss the relationship in more detail in Section~\ref{sec:rc}.

\subsection{First Attempt: Normalized PC Graphs}\label{sec:convvoting-first}

%\mred{For no space we \\ should
%	just \\ mention the first \\ two attempts?}
The first idea that comes to mind is to simply calculate the transition probabilities according to the weights on the outgoing edges from each vertex.
That is, to use the normalized PC graphs of Example~\ref{ex:normalized} and Figure~\ref{fig:firsttry}.
%for example, if from node $N$, there are two outgoing edges with weights $n_1$, $n_2$, then the probabilities associated with them could be $n_1/(n_1+n_2)$ and $n_2/(n_1+n_2)$. In the above example, this would give the graph in Figure~\ref{fig:condorcet-1}.
%In this case the normalized stationary distribution is $\frac{1}{636}(693,749,636)$. So this gives us back the intuitive order of $B$ winning, $A$ the second and $C$ the third.
Unfortunately, this does not work well.
%
%In the second example, let there be three candidates: $A$, $B$, and $C$. Let $2$ people favor $B$ to $A$, while $1$ person $A$ to $B$. Let also $2$ people favor $C$ to $B$, while $1$ person $B$ to $C$. Finally, let $3$ people favor $A$ to $C$, while no person favor $C$ to $A$.  In this case, there is no Condorcet winner. However, clearly, the strongest candidate is $A$ as it is favored the most ($1$ voters against one of the other candidates, $3$ against the other, while for $B$ and $C$ these numbers are $1$ and $2$ and $0$ and $2$) and it is disfavored the least ($0$ voters against one of the other candidates, $2$ against the other, while for $B$ and $C$ these numbers are $1$ and $2$ and $1$ and $3$).
For the election modeled in  Figures~\ref{fig:condorcet}b and  \ref{fig:firsttry}b, the stationary distribution is $(\frac{1}{2},\frac{1}{2})$ regardless of the actual numbers $n_{AB},n_{BA}$. That is, both options are ranked equally, even if almost the whole population supports $A$, which is clearly wrong. But if we consider the Markov chain in Figures~\ref{fig:firsttry}a and \ref{fig:firsttry}a, the stationary distribution is $\frac{5}{15}$ for $A$, $\frac{6}{15}$ for $B$, and $\frac{4}{15}$ for $C$. Thus, it favors $B$, while we already remarked that $A$ {seems} the strongest vertex in the corresponding Condorcet graph (Figure~\ref{fig:condorcet}a), {rather than $B$}.
%In the first, let there be two candidates, $A$ and $B$. In this case, no matter how many voters favor $A$ and how many favor $B$, this method gives probability $1$ for transition from $A$ to $B$, and also $1$ for transition from $B$ to $A$. As a result, the stationary distribution gives $1/2$ on $A$ and $1/2$ on $B$, while clearly, if more voters favor $B$, then $B$ should win. The graphs are shown in Figure~\ref{fig:firsttry}

%\begin{tikzpicture}[->,>=stealth',shorten >=1pt,auto,node distance=3cm,
%thick,main node/.style={circle,draw,font=\sffamily\Large\bfseries}]
%
%\node[main node] (a) {$A$};
%\node[main node] (b) [below left of=a] {$B$};
%\node[main node] (c) [below right of=a] {$C$};
%
%\path[every node/.style={font=\sffamily\small}]
%(a) edge node  {$2$} (b)
%edge [bend left] node {$0$} (c)
%(b) edge node  {$2$} (c)
%edge [bend left] node {$1$} (a)
%(c) edge [bend left] node {$1$} (b)
%(c) edge node  {$3$} (a)
%;
%\end{tikzpicture}
%\hspace{1cm}
%\begin{tikzpicture}[->,>=stealth',shorten >=1pt,auto,node distance=3cm,
%thick,main node/.style={circle,draw,font=\sffamily\Large\bfseries}]
%
%\node[main node] (a) {$A$};
%\node[main node] (b) [below left of=a] {$B$};
%\node[main node] (c) [below right of=a] {$C$};
%
%\path[every node/.style={font=\sffamily\small}]
%(a) edge node  {$1$} (b)
%edge [bend left] node {$0$} (c)
%(b) edge node  {$2/3$} (c)
%edge [bend left] node {$1/3$} (a)
%(c) edge [bend left] node {$1/4$} (b)
%(c) edge node  {$3/4$} (a)
%;
%\end{tikzpicture}

What is the problem?
Intuitively, the outgoing edges from each vertex are not normalized with respect to the same standards. The normalized edges always sum up to 1, regardless of whether they originate in a popular candidate or not.
%From an unpopular candidate the outgoing edges have high weight, but still, they are only compared with each other, not with outgoing edges from popular candidates.
In other words, the outgoing ``flow'' from popular candidates is exactly the same as the ``flow'' from unpopular ones, while clearly the former should be much less than the latter. The normalizing factor at each vertex is different: if the weights on the outgoing vertices are small, the factor is small, while if the weights are large, the normalizing factor is large thus distorting the original proportions.

\subsection{Solution: Adding Complement Loops}\label{sec:convvoting-final}

What can we do to normalize with the same denominator at each vertex?
We propose to introduce loops around the vertices and put as much weight on them as to complement the weights on the ``normal'' outgoing edges to the same (and large enough) number $N$.
That is, the weights on all the outgoing edges (the ``normal'' ones as well as the loops) will add up to $N$. How shall we chose $N$?  Proposition \ref{prop:anycomplete} will show that  it does not matter as long as it is large enough.
%It only remains to decide what reference value $N$ should be used to obtain the complements.
%
Specifically for Condorcet graphs, we suggest that the complemented graph is constructed by adding loops such that, for any vertex, \emph{the sum of the weights on all the outgoing edges is $|\vots| \cdot (|\cans| - 1)$}.
Notice that this procedure works for total as well as partial preferences of voters, i.e., not all the voters need to compare every pair of voting options.
{We show this formally in Section~\ref{sec:properties}.}
The construction also has some appealing common-sense interpretations that will be presented in Section~\ref{sec:interpretations}.

Finally note that when there are more than one closed communication classes {in the PC graph} (that is, when there are groups of candidates such that candidates in different groups are not compared by any voter), then the Markov stationary distribution is not unique. Hence we define our ranking to be given by the stationary distribution {obtained from the initial uniform distribution}.
%that is the limiting distribution when we start from the uniform distribution.
Here we state our final definitions rigorously:
\begin{definition}[Complemented Condorcet Graph]\label{def:ccg}
Let $\vots$ be a set of voters and let $\cans$ be a set of voting options. Let $(\pref_1,\dots,\pref_{|\vots|})$ $\in \ord^{|\vots|}$ be a preference profile. Consider the Condorcet graph  $(\vex,E)$ associated with  the preference profile. We define the \emph{complemented Condorcet graph as the pair} $(\vex,E^c)$ where
$E^c(x_1,x_2) := E(x_1,x_2)$ whenever $x_1\neq x_2$, and
$E^c(x,x):= |\vots| \cdot (|\cans| - 1) - \sum_{x'\in \vex, x' \neq x}E(x,x')$.
\end{definition}

\begin{definition}[Convergence Voting]\label{def:cv}
	Let $\vots$ be a set of voters and let $\cans$ be a set of voting options. Let ${(\pref_1,\dots,\pref_{|\vots|})} \in \ord^{|\vots|} $ be a preference profile. Consider the complemented Condorcet graph  $(\vex,E)$ associated with  the preference profile. This graph determines a Markov chain over the set of states $\vex$ taking the transition matrix $\tm := (\tme_{ij})_{i,j \in \nats}$ to be
	 $\tme_{ij}:= E(x_i,x_j) / (|\vots| \cdot (|\cans| - 1))$. We can rank the voting options $\vex = \cans$ according to the largeness of the stationary distribution of this $\tm$ given by starting from the uniform distribution.
	 We {use the term \emph{convergence voting} for} the  social choice function that assigns to each preference profile the ranking obtained this way.
\end{definition}

\begin{figure}[t]
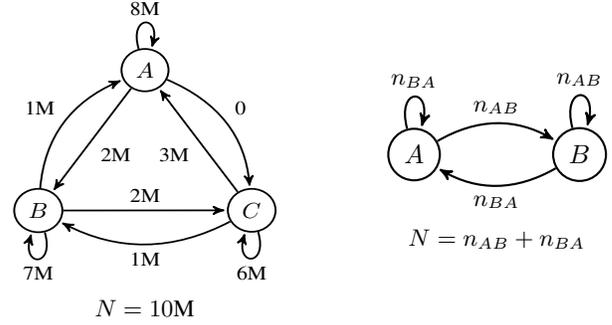
\centering
			\begin{tabular}{@{}c@{\qquad\qquad}c@{}}
		\begin{tabular}{@{}c@{}}
							\resizebox{3.5cm}{3.9cm}{%
				\begin{mytikzgraph}
					\input{graphABC-loop-compl}
				\end{mytikzgraph}}\\
			\footnotesize{$N= 10$M}
			\end{tabular}		
		&
		\begin{tabular}{@{}c@{}}
%						\resizebox{2.5cm}{1.5cm}{%
			\begin{mytikzgraph}
				\input{graphAB-loop-compl}
			\end{mytikzgraph}
%			}
		\\  \footnotesize{$N= n_{AB} + n_{BA}$}
		\end{tabular}
%		&
%		\begin{tabular}{@{}c@{}}		
%				\resizebox{3.1cm}{2cm}{%
%			\begin{mytikzgraph}
%				\input{markovchainAB-loop-compl}
%			\end{mytikzgraph}}
%		\end{tabular}
	\end{tabular}
	\caption{Complemented Condorcet graphs: (a) of Example~\ref{ex:presidential};\ (b) of Example~\ref{ex:parliamentary}}
	\label{fig:finaltry-1} \label{fig:finaltry-2}
\end{figure}

To see how it works, consider the two-party election in Example~\ref{ex:parliamentary}. The new construction produces the complemented PC graph in Figure~\ref{fig:finaltry-1}b. The normalizing factor is $n_{AB}+n_{BA}$, and the stationary distribution turns out to be $(\frac{n_{BA}}{n_{AB}+n_{BA}},\frac{n_{AB}}{n_{AB}+n_{BA}})$. Hence $A$ wins.

%Note that the procedure is well defined for total as well as partial preferences of voters, i.e., also when not all the voters compare every pair of voting options.
% However, there is an obvious one that is always good: the number of voters times the number of pairs of candidates. This is always larger than the sum of the weights of the outgoing edges.
%Once we do the completion with loops the way we suggested, the sum of the weights on all outgoing edges becomes $|\vots| \cdot (|\cans| - 1)$. When not all voters vote about each pair, then the sum of the weights on the two edges between two candidates is less than $|\vots|$. Nevertheless, we can still complete the graph with the loops such that the sum of the weights on all outgoing edges becomes $|\vots| \cdot (|\cans| - 1)$.
For the presidential election in Example~\ref{ex:presidential}, we obtain the graph in Figure~\ref{fig:finaltry-2}a, with normalizing factor $10$M and stationary distribution $( \frac{5}{11}, \frac{4}{11}, \frac{2}{11})$.
Thus, $A$ is ranked best, with $B$ not far behind, and $C$ is the weakest candidate.
%with transition matrix
%\[\fracs 1 \begin{pmatrix}
% 4 & 2 & 0\\
%1 & 3 & 2 \\
%3 & 1 & 2
% \end{pmatrix}\]

\subsection{Well-Definedness}\label{sec:properties}

We now show that the construction is well-defined: the weights are positive,  the limit always exists, and  the construction is not sensitive to the actual choice of the reference value $N$. % and robust with respect to the reference value.

\begin{proposition}\label{prop:wellformed}
	For every loop $(x,x)$ in a complemented Condorcet graph constructed according to Definition~\ref{def:ccg}, we have that $E(x,x) \ge 0$.
\end{proposition}
\begin{proof}
The  weight on each outgoing edge can be at most the number of voters $|\vots|$, and from each vertex, there can be at most $|\cans|-1$ outgoing edges, so $|\vots| \cdot (|\cans| - 1) \geq \sum_{x'\in \vex, x' \neq x}E(x,x')$ for all $x \in \cans$
\end{proof}

\begin{proposition}
The Markov process of the complemented Condorcet graph is such that the irreducible components are all aperiodic, hence the limit exists (by Section \ref{sec:markov}).
\end{proposition}
\begin{proof}
	A state  belonging to an irreducible component must have an incoming edge with non-zero weight by definition of  irreducibility. Then it also has a loop if the normalizing factor is at least
	$|\vots| \cdot (|\cans| - 1)$, and aperiodicity follows.
	\end{proof}

\begin{proposition}\label{prop:anycomplete}
	The outcome of Convergence Voting is the same if we replace the normalizing factor $|\vots| \cdot (|\cans| - 1)$ with any larger $N$. 	
\end{proposition}
\begin{proof} If the transition matrix is $\tm$, the
	 effect of increasing the normalizing factor is to
	  switch to $\tm' = r\tm + (1-r){\bf I}$ for some
	   $r\in (0,1)$. The eigenvectors with
	    eigenvalue $1$ are clearly the same for $\tm'$  and
	    $\tm$. Working with the canonical forms, the irreducible components have
	     only a single stable distribution each, so
	      $\tm^\infty_k$ and $\tm^{\prime \infty}_k$
	        agree. Also, ${\bf
	     S}^{\prime\infty}_k = ({\bf I} - {\bf Q}')^{-1} {\bf S}'_k \tm^{\prime\infty}_k =
      ({\bf I} - (r{\bf Q} + (1-r){\bf I}))^{-1} r{\bf S}_k \tm^\infty_k =
      ({\bf I} - {\bf Q})^{-1} {\bf S}_k \tm^\infty_k = {\bf S}^{\infty}_k$. Hence $\tm^\infty = \tm^{\prime\infty}$.
\end{proof}

\subsection{Graph-Theoretic Interpretation and Computational Properties}\label{sec:computability}

The complemented Condorcet graph in Definition~\ref{def:ccg} has the following graph-theoretic interpretation.
For each ordered pair of voting options $(A,B)$, we construct an \emph{atomic complemented PC graph $G_{AB}$} with vertices $\{A,B\}$ and edges $\{(A,B), (A,A)\}$. The edge $(A,B)$ from $A$ to $B$ is labeled by the number of voters who prefer $B$ over $A$. The loop $(A,A)$ is labeled by the number of the remaining voters.
Thus, $G_{AB}$ is focused solely on who prefers $B$ over $A$, and who does not.
This can be formalized as follows.

\begin{definition}[Union of graphs]
	Given two (not necessarily disjoint) graphs $(\vex_1,E_1)$ and $(\vex_2,E_2)$, their \emph{union} is obtained by taking the union of their vertices, and summing up the weights on the edges.
	Formally, it is defined as $(\vex,E)$ with $\vex = \vex_1\cup \vex_2$ and $E(x,y) = \hat{E}_1(x,y) + \hat{E}_2(x,y)$, where $\hat{E}_i(x,y) = E_i(x,y)$ whenever $x,y\in \vex_i$ and $0$ otherwise.
\end{definition}

\begin{proposition}%[Interpretation of the graphs]
The complemented Condorcet graph is the union of the atomic complemented PC graphs for all the pairs from $\cans$.
\end{proposition}
\begin{proof}
Straightforward.
\end{proof}

Finally, the ranking has good computational properties.

\begin{proposition}\label{prop:complexity}
The ranking can be computed in {deterministic polynomial time}, more precisely in {$O(|\cans|^2\cdot|\vots| + |\cans|^3)$}.
\end{proposition}
\begin{proof}
{The Condorcet graph and the corresponding Markov chain can be prepared in time $O(|\cans|^2\cdot|\vots|)$.
Then, the computation of the stationary distribution} only involves inverting matrices of at most $|\cans|$ size, which is known to be $O(|\cans|^3)$ for Gauss elimination.
%By the properties of discrete-time, time-homogeneous finite Markov chain, presented in Section~\ref{sec:markov}, we have that the scores can be computed by solving a system of $k$ linear equations with $k$ variables and rational coefficients, where $k$ is the number of candidates.
\end{proof}

{It is important to notice that the number of voting options $|\cans|$ is typically small, and while the number of voters $|\vots|$ can be huge, the computation is linear with respect to $|\vots|$.}

\subsection{How to Use the Ranking}\label{sec:election-outcome}

%So far, we have focused on how to produce a ranking of the voting options, based on pairwise comparisons.
The ranking can be used to determine the winner(s) at least in two meaningful ways.
First, for a single-winner vote, the winning option can be defined as the option with the highest score.

\begin{example}
Consider the presidential election scenario of Example~\ref{ex:presidential}, which produces the stationary distribution of $(5/11, 4/11, 2/11)$ for $A$, $B$, and $C$ respectively.
According to the above interpretation, $A$ becomes the winner with the highest score of $5/11$.
\end{example}

	%\item[Randomize:]
	%Select the winner at random, according to the stationary distribution. In the example, this means choosing $A$ with probability $0.455$, $B$ with probability $0.364$, and $C$ with probability $0.182$.

Secondly, for elections of representative bodies (such as a parliament, university council, etc.), the seats can be divided between the parties according to the stationary distribution.

\begin{example}
Take the scenario in Example~\ref{ex:presidential}, and assume now that it corresponds to a parliamentary election involving parties $A$, $B$, and $C$.
For the distribution of $(5/11, 4/11, 2/11)$, and a 110-seat assembly, the simplest choice is to allocate $50$ seats to party $A$, $40$ seats to $B$, and $20$ to $C$.
\end{example}

We note that the latter approach can be easily combined with the existing practical methods of seat allocation, such as the Jefferson/D'Hondt method~\cite{Gallagher91proportionality}.

%\brown{Note also that the last approach can be easily combined with the existing practical methods of seat allocation. For instance, applying the Jefferson/D'Hondt \red{What is that?} method we obtain the output in Figure~\ref{fig:dhondt}, and the parties get 7, 5, and 2 seats, respectively.
%\begin{figure}[t]\centering
%	\includegraphics[width=\columnwidth]{dhondt.pdf}
%	\caption{\red{Would be good to remove yellow} Table for Jefferson/D'Hondt. The columns show the output for subsequent denominators. The entries corresponding to an allocated seat are highlighted in bold}
%	\label{fig:dhondt}
%\end{figure}}

\section{Convergence Voting as Consensus Reached by Negotiation}\label{sec:interpretations}

In this section, we present two interpretations of the ranking, which give justification to our voting rule.
The first, proposed in Sections~\ref{sec:ncs}--\ref{sec:general-ncs},  interprets the graph in terms of arriving at an agreed community support through an  infinite sequence of imaginary negotiating steps.
The second, in Section~\ref{sec:icd}, explains the convergence graph by means of iterated probabilistic change of the decision under consideration.

It is important to observe that the final score of a voting option in our voting represents a quantification of the voters' community's support to be the most preferred option. If there is a single voter, the end score will allocate $1$ to their highest preference and $0$ to the others. But then one may ask why not just count how many times each voting option is preferred the most among the voters?  The reason is that that method entirely ignores whether a voting option is the second preference or
the least preferred option of a voter for whom it is not the most preferred option. One way to take that into account is the Borda count, to which we compare our method in Section %\ref{sec:ncs} and
\ref{sec:Borda}.

%\NB{What about non-condorcet PC graphs?}

\subsection{Negotiated Community Support}
\label{sec:ncs}

Assume a set of possible collective decisions of a community (i.e., voting options $\cans$), together with a representation of pairwise preferences in the form of the Condorcet graph.
Suppose further that we want to quantify the aggregate support for each voting option within the community. The following structure can be useful.
%(such as for allocating resources).

\begin{definition}[Support Function]
Let $\cans$ be the set of voting options. We call a function $s: \cans \rightarrow [0,1]\subset \reals$ with $\sum_{c\in \cans}s(c) = 1$ a normalized support function.
\end{definition}

%Suppose we have a community of voters $\vots$, and we want to find the normalized support-function $s_\vots: \cans \rightarrow \reals$ that reflects the voters' aggregate support for the voting options.
The big question is how to construct an appropriate support function for an actual voting input.
If we had the individual support functions for all the voters, we could define the collective support as their average. Unfortunately, the exact support of each voter for each voting option is usually not known, and may not even exist.
%For example, if each voter $v\in \vots$ has a  support function $s_v$, then simply by taking the normalized sum (average) of all individual support functions, we obtain an aggregate support function $s_\vots : \frac{1}{|\vots|}\sum_{v\in \vots} s_v$. However, normally individuals do not have such support functions or it is not known.
%
What we have instead is a preference list from each voter $v$, i.e., a strict partial order
$o_v$ on $\cans$, {or just pairwise comparisons}.
%In Condorcet voting, it is usual that the voters have to provide a preference list. So let $\ord$ be the set of strict partial orders on $\cans$,
%(equality could also be allowed, e.g. $c_1 > c_2 = c_3 = c_4 > c_5 = c_6\ , \ c_2> c_7=c_8$ etc.),
%and suppose that for each voter $v$, an ordering $o_v \in \ord$ is given.
A simple way to compute aggregate support from the $o_v$'s is given by the Borda count: {for each voter $v$,} assign the highest rank of $o_v$ with the score $|\cans|$, the next one with $|\cans|-1$ etc, then add the scores for each option {as $v$ runs through all voters}, and normalize.
However, creating the individual support functions this way is somewhat ad hoc.
In particular, it assumes that the difference between two subsequent options in $o_v$ is always the same, which is in general hard to justify.
%$o_v$ is just an ordering it does not quantify how much the voter prefer one option over another; the Borda method is an ad hoc way to quantify the support by assigning a unit value to the difference between an option and the next preferred option.
%\red{(WJ: I would remove the following sentence - it seems a weak argument to me, as nobody forces us to do Condorcet voting.)
{Furthermore, it may actually be easier to just collect pairwise comparisons, like in a machine learning setting.
}
% for each pair of candidates $c, c'\in \cans$, the number of voters preferring $c$ to $c'$, and those preferring $c'$ to $c$ are recorded. Let $E (c,c') \in \nats$ be this number.

We propose the following procedure to model how a group of voters arrives at a reasonable community support function given a Condorcet graph:
\begin{itemize}
\item Start with a uniform support function $s^0$ on $\cans$.
\item The community members carry out a sequence of imaginary negotiation steps, through which they modify the support function and obtain $s^1$, $s^2$, etc.
\item If this sequence converges, then the limiting function expresses the aggregate support.
\end{itemize}
Moreover, we can define the negotiation process as follows.
\begin{enumerate}
\item At each step $s^i$, each voter $v\in \vots$ gets a share $\frac{1}{|\vots|}s^i$ that she rearranges according to the following rule:
\item For each voting option $c$, the voter divides her share $\frac{1}{|\vots|}s^i(c)$ for that candidate into $(|\cans|-1)$ parts corresponding to the other voting options $c' \in \cans \setminus\{c\}$;
\item The part $\frac{1}{|\vots|\cdot(|\cans|-1)}s^i(c)$ corresponding to $c' \in \cans \setminus\{c\}$ is transferred from $c$ to $c'$ if $v$ prefers $c'$ over $c$, otherwise it is kept at $c$;
\item Then, the result of all transfers for each voter are added up, obtaining $s^{i+1}$.
\end{enumerate}

%Since this is the  community's support, at each step $s^i$, each voter $v\in \vots$ can have a share $\frac{1}{|\vots|}s^i $ that he can rearrange according to some rule. What rule?
%One reasonable rule is that for each voting option $c$, the voter further divides their share $\frac{1}{|\vots|}s^i (c)$ into $|\cans|-1$ parts corresponding to the $|\cans|-1$ other voting options: $\cans \setminus\{c\}$.
%The part $\frac{1}{|\vots|\cdot(|\cans|-1)}s^i (c)$ corresponding to $c' \in \cans \setminus\{c\}$ is then kept at $c$ if $v$ prefers $c$ to $c'$ or if $v$ is indifferent between these options, and it is transferred to $c'$ if $v$ prefers $c'$ to $c$ (we could also allow equality: in that case it is equally distributed between $c$ and $c'$ if $v$ lists them as equal). Then the result of all transfers for each voter are added up, obtaining $s^{i+1}$.

%If we look at $s^{i+1}$ as a Markov chain, then this Markov chain has exactly the same transition matrix as the Markov chain we defined in the previous chapter. By Markov's theory, $s^{i}$ converges to a stationary distribution, which we take to be $s_\cans$. This $s_\cans$ is then stationary under re-negotiation, and we can say that it represents an agreement in the community reached by this imaginary process.

\noindent
The following is immediate from the definitions.
\begin{proposition}
	For every $i=0,1,\dots$, the process $s^{i}$ is a Markov chain with the same transition matrix {as our convergence voting transition matrix in Definition \ref{def:cv}}.
%defined in the previous section.
\end{proposition}
By Markov's theory, the sequence $s^{0}, s^{1}, \dots$ converges to a stationary distribution, $s_\cans$. This $s_\cans$ is stationary under re-negotiation {by the above rules}, and {hence}
%we
 it represents an agreement in the community reached by this imaginary process.

\subsection{Generalizing the Negotiation Process}
\label{sec:general-ncs}

We can of course consider other ways of rearranging the individual support in each turn of the negotiations. For example, voters could rearrange all their support to their most preferred candidate, leaving nothing for the others. In this case the resulting stationary distribution of the Markov chain ranks the candidates according to the number of voters who give them the highest rank.
We could also imagine that voters rearrange their support according to their individual quantified support. In that case the stationary distribution is the same as simply computing the normalized sum of the individual scores.

This idea can be easily generalized. Each voter  $v$'s negotiating position could be represented by a transition matrix $\tm_v$,
all rows adding to $1$.
The component $(\tm_v)_{ij}$ represents the fraction of $v$'s share in the current community support of the $i$'th voting option that $v$ would rather transfer to $j$. Clearly, $\sum_{i \in \cans} (\tm_v)_{ij} =1$ must hold.
Such a matrix can be called $v$'s \emph{negotiating position for support redistribution}.
In the above special case
for our convergence voting $(\tm_v)_{ij} =1/(|\cans|-1)$ if $v$ prefers $j$ to $i$, otherwise $0$, and $(\tm_v)_{ii}$ is what remains from $1$.  {Another special case is when voter $v$ knows exactly how they  want distribute their share of resources (i.e. $v$ has his individual support function) then $\tm_v$ has identical
rows agreeing with $v$'s support function}.  If the voters are not equal (for example, their share of support could be weighted by the fraction of stock they hold in a company), then to each voter $v$ a share $h_v$ is assigned such that  $\sum_{v\in \vots} h_v =1$.
Then the weighted sum $\tm = \sum_{v\in \vots} h_v \tm_v$ gives the transition matrix that rearranges all the
supports of the voters according to their wish and their share of the support.
In consequence, we arrive at the following definition.
\begin{definition}[Renegotiated Community Support]
Let $\vots$ be a community of voters, let $\cans$ be a set of voting options. Let $s$ be an initial normalized support function on $\vots$. Let the transition matrix $\tm_v$ denote voter $v$'s negotiating position for the support rearrangement.
Let $h_v$ be voter $v$'s share in the support. Then $\tm := \sum_{v\in \vots} h_v \tm_v$ defines a transition matrix for a
Markov chain over $\cans$, and the stationary distribution $s'$ that is reached starting from $s$ is called \emph{renegotiated normalized community support}. If there is no initial support function, uniform distribution is used as the initial support, {and we can call it \emph{negotiated normalized community support}}.
\end{definition}

%\subsection{Election Outcome as Random Walk}\label{sec:graph-distribution}
%
%\red{Recall the property mentioned by Gergei (the probability that a random walk will feature the frequencies from the stationary distributions is equal to 1).
%	Thus, if the community reallocates the support according to the EPC graph (or, more precisely, its Markov chain), then the stationary distribution indicates, for each voting option, how often the option will be selected in the long run.}

\subsection{Iterated Change of Decision}\label{sec:icd}

% DN: I was always worried by the insertion of randomness into the interpretation of the graph to get the result out. One may argue that a given same graph with two different random starts gives different outcomes. To resolve this I suggest that once the graph is ready, the graph should be hashed to provide the random tape allowing to perform the random walk. That way we can argue that as long as the graph is not known in advance there is ground to assume that it would be very hard to select a "proper" random to influence the interpretation of the vote.

To construct the second interpretation of convergence voting, we observe that
the transition matrix given in Definition~\ref{def:cv} corresponds to the following random process.
%At vertex $x$:
\begin{enumerate}
\item Select a {uniformly} random voting option $c\in\cans$ as the tentative output.
\item Repeat:
\begin{enumerate}
\item choose uniformly an alternative voting option $c'\neq c$;
\item choose uniformly a voter $v\in\vots$;
\item if $v$ prefers $c'$ to $c$, then change the tentative output to $c'$, else stay with $c$.
% If $V$ ranks $Y$ and $X$ as equal, then proceed to $Y$ or stay with one half - one half probability.
% WJ: We don't have weak preferences for the moment!
\end{enumerate}
\end{enumerate}

%\red{Wojtek, could you please elaborate on the interpretation below? Currently it is very unclear. How is this a deliberation process. What is "contemplating" a candidate? How is this related to the limiting distribution?}
Thus, one can interpret the Markov chain in Definition~\ref{def:cv} as a specification of an infinite iterated process of collective deliberation. Each round produces a tentative collective decision. At the next round, the community puts forward an alternative and asks a random member if this alternative is better than the current decision.
%shifting the preference of the comunity from one voting option to another.
The shift is done according to the member's pairwise preferences.
%I.e., if the population currently prefers option $x$, the option to be preferred in the next round is given by the random choice of the above process. Namely, in each round, the next preferred option is decided by randomly choosing a voter, then randomly choosing another option, and if the new option is preferred by the chosen voter, the preference is shifted to the new option, otherwise remains the same.

The voting options can be assigned intermediate scores at each step, defined by the frequency of being the preferred decision until that time.
By the ergodic theorem, with probability $1$, the sequence of intermediate scores on each option for an infinite number of iterations converges to our score  given by the stationary distribution to that option.

\section{Further Properties}
\label{sec:further-properties}

In this section, we look at the classical properties of Arrow's Theorem~\cite{Arrow50impossibility}. We also discuss the case of Condorcet graphs that are not strongly connected, i.e., consist of multiple closed communication classes.

\subsection{Arrow's Properties}

\begin{theorem}[Pareto Efficiency]
	Suppose that individual voters' preferences are consistent in the sense that if a voter prefers $A$ to $B$ and $B$ to $A$ then he or she also prefers $A$ to $C$. With this condition,
	our ranking is Pareto efficient: If candidate $A$ is preferred to candidate $B$ by all voters, then $A$ is ranked higher than $B$.
	\end{theorem}
\begin{proof}
	Suppose all voters prefer $A$ to $B$. Then, given our consistency assumption, for any third candidate $C$,
	if a voter prefers $B$ to $C$, then he or she also prefers $A$ to $C$. Hence,  the weight increase on the arrow pointing from $C$ to $B$ corresponding  to this voter's preference of $B$ over $C$, comes with another weight increase on the arrow pointing from $C$ to $A$ corresponding  to this voter's preference of $A$ over $C$. Consequently, the weight on the arrow pointing from $C$ to $A$ is at least as large  as the weight pointing from $C$ to $B$. With a similar argument, we obtain that the weight on the arrow pointing from $A$ to $C$ is at most as large as the weight on the arrow pointing from $B$ to $C$, and consequently the weight on the loop around $A$ must be at least as large as  the weight on the loop around $B$. In equilibrium the inflow and the outflow to each candidate have to be the same. However,  because of the above, from each other candidate $C$, the inflow to $A$ is at least as large as the inflow to $B$.
	On the other hand, from $A$ there is no flow to $B$, so the total inflow  to $A$ is strictly larger than to $B$. On the other hand, the weights on the edges pointing away from $A$ are strictly smaller than the weights on the edges pointing away from $B$. Equilibrium then is only possible if the weight on $A$ is larger than the weight on $B$.
	\end{proof}

\begin{proposition}[No Dictatorship]
	Our ranking does not allow dictators.
\end{proposition}
\begin{proof}
	This is trivial. Any preference $A>B$ of any voter can be overturned by two other voters with preferences $A<B$.
\end{proof}
\begin{proposition}
	Our ranking is not independent of irrelevant alternatives.
\end{proposition}
\begin{proof}
To see this, consider the voting scenario in Section~\ref{sec:cwnb}. If we remove candidate $A$, then $C$ becomes the winner, while by adding $A$, who is much less popular than either $B$ or $C$ and ends up at the bottom, $B$ comes out as the winner.
\end{proof}

Note that, although the independence principle is listed by Arrow as a desirable property, this example suggests it is rather debatable. Without $A$, we only have evidence that $11$ voters prefer $C$ and $9$ prefer $B$. But we do not have evidence about how much more those $11$ voters prefer $C$ to $B$ and vice versa. Perhaps those $11$ voters prefer $C$ slightly, but the $9$ prefer $B$ very much over $C$. When we add $A$, that brings additional evidence to the table about the preferences. Namely, it shows that in fact much more voters prefer $B$ to $A$ then $C$ to $A$. In other words, relative to $A$, $B$ is much stronger than $C$, so much so, that this helps $B$ overcome his or her weakness when compared directly with $C$. This is what we obtained by looking at the preference lists as well: $4$ voters seem to have a wider gap between $B$ and $C$ fitting $A$ in between.

We also note that our method is not monotonic. Monotonic would mean that for a given preference list and
winner $X$, if we modify the list such that for each voter, whichever voting option is favored less than $X$ is
kept favored less, then the resulting profile still results in the same winner $X$.
However, if we modify the preference list in Section \ref{sec:cwnb} to $A\pref C \pref B : 8$ voters;\
$C\pref A \pref B : 0$ voters, then $B$ is not the winner any more, but our method also delivers the Condorcet winner $C$.

\subsection{Multiple Closed Communication Classes}

As we mentioned in Sections \ref{sec:markov} and \ref{sec:convvoting-final}, although the limit distribution when we start from the uniform distribution always exists and is unique, it may not be a unique stationary distribution.
That is the case when in the Markov chain there are more than one closed communication classes. That happens when there are at least two groups of voting options $\cans_1, \cans_2 \subset \cans$ such that no voter compares any option in $\cans_1$ with any option in $\cans_2$. Even in this case, the limiting distribution starting from the uniform distribution will give a reasonable ranking. But it is important that the voters  understand that not comparing two options means not caring about which gets higher support. Even if just one voter prefers  an option in group $\cans_1$ to an option in group $\cans_2$, the limit distribution  will end up  entirely on $\cans_1$.
While in a large-scale election this is highly unlikely, in the case when there are few voters, it is important that the voters understand this possibility.
If this is undesirable, there are options to avoid it. For example, it can be postulated that votes between pairs are only entered in a graph if they reach a certain percentage. Or, our voting can be somewhat modified to allow only a single list to be submitted by a voter and unlisted options are taken into account equally at the bottom. (From Section \ref{sec:ncs}, it is clear that our technique can be extended to allow equality in voter listing, which is then entered in the graph  with $1/2$ weights on both arrows between the two candidates in question.)

%\section{Things to Do}

%WJ: define the pathological outcomes based on ``incidental dictatorship'' (if we remove 1 voter, or a small subset of voters, then it changes the winner of the election, and it *does* matter which voter we remove).
%(Note: I suspect that Copeland on partial preferences also admits such pathological behavior!).

%More generally: small perturbation in the input leads to a large change in the output, and symmetric perturbations lead to completely different changes.

\section{Comparison to Existing Ranking Methods}
\label{sec:comparison}

Here, we compare convergence voting to relevant ranking methods that try to aggregate the values of options in a balanced way.

\subsection{MC3/Rank Centrality}\label{sec:rc}

A similar ranking based on pairwise comparisons has been suggested in a different context  as a statistical estimator first by Dwork et al. \cite{Dwork01mc3} called MC3, and then
by Negahban et al. \cite{Negahban12rankcentrality} calling it Rank Centrality. Their ranking coincides with ours  when each pair of candidates is either compared by all voters or none.
For partial preferences, the two rankings differ significantly.
In MC3/Rank Centrality only the voters who compare $x$ and $y$ are chosen uniformly in the view of Section \ref{sec:icd}.
As a result, the transitions for MC3/Rank Centrality do not depend on how many voters compared a given pair, only their preference ratios.

\begin{figure}[t]
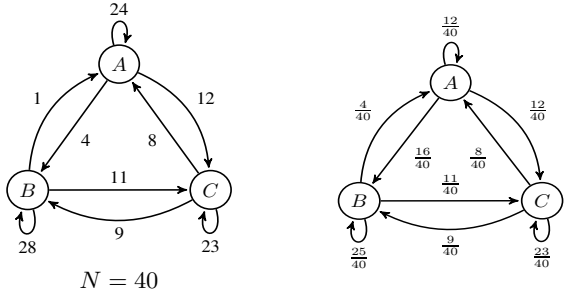
\centering
	\begin{tabular}{@{}c@{\qquad\qquad}c@{}}
		\begin{tabular}{@{}c@{}}
			\resizebox{3cm}{3.5cm}{%
				\begin{mytikzgraph}
	\node[main node] (a) {$A$};
\node[main node] (b) [belowleft] {$B$};
\node[main node] (c) [belowright] {$C$};

\path[every node/.style={font=\footnotesize}]
(a) edge node  {4} (b)
edge [bendleft] node {12} (c)
edge [loop above] node {24} (a)
(b) edge node  {11} (c)
edge [bendleft] node {1} (a)
edge [loop below] node {28} (b)
(c) edge [bendleft] node {9} (b)
edge node  {8} (a)
edge [loop below] node {23} (c)
;
			\end{mytikzgraph}}\\
			\footnotesize{$N= 40$}
		\end{tabular}		
		&
		\begin{tabular}{@{}c@{}}
			\resizebox{3cm}{3.5cm}{%
				\begin{mytikzgraph}
						\node[main node] (a) {$A$};
\node[main node] (b) [belowleft] {$B$};
\node[main node] (c) [belowright] {$C$};

\path[every node/.style={font=\footnotesize}]
(a) edge node  {$\frac{16}{40}$} (b)
edge [bendleft] node {$\frac{12}{40}$} (c)
edge [loop above] node {$\frac{12}{40}$} (a)
(b) edge node  {$\frac{11}{40}$} (c)
edge [bendleft] node {$\frac{4}{40}$} (a)
edge [loop below] node {$\frac{25}{40}$} (b)
(c) edge [bendleft] node {$\frac{9}{40}$} (b)
edge node  {$\frac{8}{40}$} (a)
edge [loop below] node {$\frac{23}{40}$} (c)
;
			\end{mytikzgraph}}
		\end{tabular}
	\end{tabular}
	\caption{Comparison to MC3/Rank Centrality:\ (a) Complemented Condorcet graph;\ (b) Markov chain produced by MC3/Rank Centrality}
	\label{fig:rankcen}
\end{figure}

Consider for example the complemented Condorcet graph in Figure~\ref{fig:rankcen}a, with the number of votes on the edges between candidates. We obtain the convergence voting probabilities simply by normalizing with $N= 40$.
\ignore{

\begin{tikzpicture}[->,>=stealth',shorten >=1pt,auto,node distance=3cm,
thick,main node/.style={circle,draw,font=\sffamily\Large\bfseries}]

\node[main node] (a) {$A$};
\node[main node] (b) [below left of=a] {$B$};
\node[main node] (c) [below right of=a] {$C$};

\path[every node/.style={font=\sffamily\small}]
(a) edge node  {$4$} (b)
edge [bend left] node {$12$} (c)
%edge [loop above] node {$24$} (a)
(b) edge node  {$11$} (c)
edge [bend left] node {$1$} (a)
%edge [loop left] node {$29$} (b)
(c) edge [bend left] node {$9$} (b)
edge node  {$8$} (a)
% edge [loop right] node {$22$} (c)
;
\end{tikzpicture}
\begin{tikzpicture}[->,>=stealth',shorten >=1pt,auto,node distance=3cm,
thick,main node/.style={circle,draw,font=\sffamily\Large\bfseries}]

\node[main node] (a) {$A$};
\node[main node] (b) [below left of=a] {$B$};
\node[main node] (c) [below right of=a] {$C$};

\path[every node/.style={font=\sffamily\small}]
(a) edge node  {$\frac{4}{40}$} (b)
edge [bend left] node {$\frac{12}{40}$} (c)
edge [loop above] node {$\frac{24}{40}$} (a)
(b) edge node  {$\frac{11}{40}$} (c)
edge [bend left] node {$\frac{1}{40}$} (a)
edge [loop left] node {$\frac{28}{40}$} (b)
(c) edge [bend left] node {$\frac{9}{40}$} (b)
edge node  {$\frac{8}{40}$} (a)
edge [loop right] node {$\frac{23}{40}$} (c)
;
\end{tikzpicture}
}
Our aggregate  ranking in this case turns out to be $A \pref B \pref C$, as it should. Note that $C$ is also the the Condorcet winner: $C$ is preferred to both $A$ and $B$, and the
flow resulting from the few votes between $A$ and $B$ cannot overturn this. If we follow the
MC3/Rank Centrality definition, then the transition probabilities  can be seen in Figure~\ref{fig:rankcen}b.
\ignore{\begin{tikzpicture}[->,>=stealth',shorten >=1pt,auto,node distance=3cm,
thick,main node/.style={circle,draw,font=\sffamily\Large\bfseries}]

\node[main node] (a) {$A$};
\node[main node] (b) [below left of=a] {$B$};
\node[main node] (c) [below right of=a] {$C$};

\path[every node/.style={font=\sffamily\small}]
(a) edge node  {$\frac{16}{40}$} (b)
edge [bend left] node {$\frac{12}{40}$} (c)
edge [loop above] node {$\frac{12}{40}$} (a)
(b) edge node  {$\frac{11}{40}$} (c)
edge [bend left] node {$\frac{4}{40}$} (a)
edge [loop left] node {$\frac{25}{40}$} (b)
(c) edge [bend left] node {$\frac{9}{40}$} (b)
edge node  {$\frac{8}{40}$} (a)
edge [loop right] node {$\frac{23}{40}$} (c)
;
\end{tikzpicture}}
This Markov chain results in the ranking $A \pref C \pref B$. That is, the strong flow from $A$ to $B$ overturns the
advantage of $C$ in favor of $B$. While  we argue in the next subsection that this is reasonable when the
votes between $A$ and $B$ are $4:16$, it seems unreasonable to allow any small voter participation between $A$ and $B$ to have the same effect.

\subsection{Condorcet and Copeland}\label{sec:cwnb}

\begin{figure}[t]
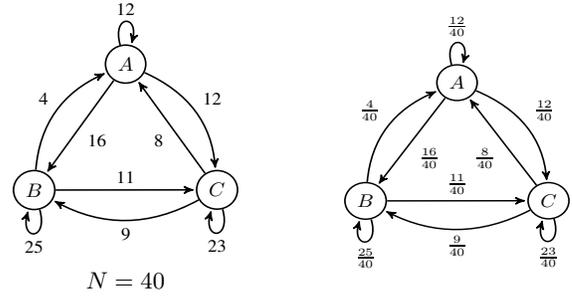
\centering
	\begin{tabular}{@{}c@{\qquad\qquad}c@{}}
		\begin{tabular}{@{}c@{}}
			\resizebox{3cm}{3.5cm}{%
				\begin{mytikzgraph}
	\node[main node] (a) {$A$};
\node[main node] (b) [belowleft] {$B$};
\node[main node] (c) [belowright] {$C$};

\path[every node/.style={font=\footnotesize}]
(a) edge node  {16} (b)
edge [bendleft] node {12} (c)
edge [loop above] node {12} (a)
(b) edge node  {11} (c)
edge [bendleft] node {4} (a)
edge [loop below] node {25} (b)
(c) edge [bendleft] node {9} (b)
edge node  {8} (a)
edge [loop below] node {23} (c)
;
			\end{mytikzgraph}}\\
			\footnotesize{$N= 40$}
		\end{tabular}		
		&
		\begin{tabular}{@{}c@{}}
			\resizebox{3cm}{3.5cm}{%
				\begin{mytikzgraph}
						\node[main node] (a) {$A$};
\node[main node] (b) [belowleft] {$B$};
\node[main node] (c) [belowright] {$C$};

\path[every node/.style={font=\footnotesize}]
(a) edge node  {$\frac{16}{40}$} (b)
edge [bendleft] node {$\frac{12}{40}$} (c)
edge [loop above] node {$\frac{12}{40}$} (a)
(b) edge node  {$\frac{11}{40}$} (c)
edge [bendleft] node {$\frac{4}{40}$} (a)
edge [loop below] node {$\frac{25}{40}$} (b)
(c) edge [bendleft] node {$\frac{9}{40}$} (b)
edge node  {$\frac{8}{40}$} (a)
edge [loop below] node {$\frac{23}{40}$} (c)
;
			\end{mytikzgraph}}
		\end{tabular}
	\end{tabular}
	\caption{(a) Complemented Condorcet graph;\ (b) Markov chain produced by our method}
	\label{fig:notcondor}
\end{figure}

It is easy to see that our function does not have to select the Condorcet winner even if one exists.
Consider the following set of preference lists:\
$A\pref C \pref B : 4$ voters;\
$C\pref A \pref B : 4$ voters;\
$C \pref B \pref A: 1$ voters;\
$A\pref B \pref C : 8$ voters;\
$B\pref A \pref C : 0$ voters;\
$B\pref C \pref A: 3$ voters.
The Condorcet winner is $C$. The graphs are shown in Figure \ref{fig:notcondor}.
Our Markov chain in this case is the same as the Markov chain produced by
rank centrality in the example of Section \ref{sec:rc}. So in this case, our method also returns the aggregate ranking $A\pref C \pref B$.

We argue that this is a reasonable output, and that in fact the Condorcet winner is not always the best choice.
While the margin of $C$ against $B$ is just two votes, and against $A$ it is $4$, the margin of $B$ against $A$ is $12$. This large margin gives  $B$ an advantage over $C$.
In this case, in the $C\pref B$ relation, for $5$ voters the distance between $C$ and $B$  is $1$, whereas for $4$ voters it is $2$, while in the $B\pref C$ relation, the distance is always only $1$.

The above implies also that convergence voting may produce different winners than Copeland method.
%Indeed, the Borda count also gives $B$ as the top candidate.

%\brown{Even if we look at the above voting diagram, and ask, how do $B$ and $C$ compare, clearly, the answer is not
%$9:11$, because besides the direct comparison, there is the route through $A$. There, there is
%$4+12$ in favor of $C$, and $8+16$ in favor of $B$. This should also weigh in besides $9:11$.
%The question is however, how to weigh the different routes when we want to compare $B$ and $C$. We will come back to this line of thought later, to
%give another justification to our social choice function.}

\subsection{Borda}
\label{sec:Borda}
Convergence voting differs from Borda as well. %
Since not giving preferences between certain voting options has different meaning in our case (the voter does not care) and Borda (not listed options are at the bottom), we consider an example where all options are ranked by all voters.
Let us modify the example of Section \ref{sec:cwnb} as follows:\
$A\pref C \pref B : 4$ voters;\
$C\pref A \pref B : 3$ voters;\
$C \pref B \pref A: 2$ voters;\
$A\pref B \pref C : 8$ voters;\
$B\pref A \pref C : 0$ voters;\
$B\pref C \pref A: 3$ voters.
Then the Condorcet winner is still $C$, Borda still gives $B$ as the winner, whereas convergence voting produces the ranking $A\pref B \pref C$. In other words, our method takes into account the evidence that A brings to the table (as we saw in the previous example), but not as much as the Borda count does.
\begin{figure}[t]
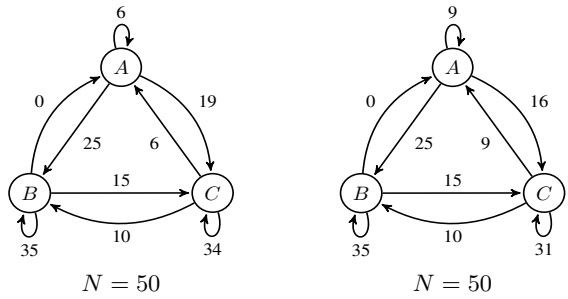
\centering
	\begin{tabular}{@{}c@{\qquad\qquad}c@{}}
		\begin{tabular}{@{}c@{}}
			\resizebox{3cm}{3.5cm}{%
				\begin{mytikzgraph}
					\node[main node] (a) {$A$};
					\node[main node] (b) [belowleft] {$B$};
					\node[main node] (c) [belowright] {$C$};
					\path[every node/.style={font=\footnotesize}]
					(a) edge node  {25} (b)
					edge [bendleft] node {19} (c)
					edge [loop above] node {6} (a)
					(b) edge node  {15} (c)
					edge [bendleft] node {0} (a)
					edge [loop below] node {35} (b)
					(c) edge [bendleft] node {10} (b)
					edge node  {6} (a)
					edge [loop below] node {34} (c)
					;
			\end{mytikzgraph}}\\
					\footnotesize{$N= 50$}
		\end{tabular}		
		&
		\begin{tabular}{@{}c@{}}
			\resizebox{3cm}{3.5cm}{%
				\begin{mytikzgraph}
					\node[main node] (a) {$A$};
					\node[main node] (b) [belowleft] {$B$};
					\node[main node] (c) [belowright] {$C$};
					\path[every node/.style={font=\footnotesize}]
					(a) edge node  {25} (b)
					edge [bendleft] node {16} (c)
					edge [loop above] node {9} (a)
					(b) edge node  {15} (c)
					edge [bendleft] node {0} (a)
					edge [loop below] node {35} (b)
					(c) edge [bendleft] node {10} (b)
					edge node  {9} (a)
					edge [loop below] node {31} (c)
					;
			\end{mytikzgraph}}\\
					\footnotesize{$N= 50$}
		\end{tabular}
	\end{tabular}
	\caption{Examples in Section~\ref{sec:Borda}. In both cases the Condorcet winner is $C$, and the Borda winner is $B$. Our method still delivers $C$ on the left, but $B$ on the right.}
	\label{fig:borda}
\end{figure}
% end-brown
%
%
\ignore{
\begin{tikzpicture}[->,>=stealth',shorten >=1pt,auto,node distance=3cm,
thick,main node/.style={circle,draw,font=\sffamily\Large\bfseries}]

\node[main node] (a) {$A$};
\node[main node] (b) [below left of=a] {$B$};
\node[main node] (c) [below right of=a] {$C$};

\path[every node/.style={font=\sffamily\small}]
(a) edge node  {$15$} (b)
edge [bend left] node {$12$} (c)
%edge [loop above] node {$24$} (a)
(b) edge node  {$11$} (c)
edge [bend left] node {$5$} (a)
%edge [loop left] node {$29$} (b)
(c) edge [bend left] node {$9$} (b)
edge node  {$8$} (a)
% edge [loop right] node {$22$} (c)
;
\end{tikzpicture}
\begin{tikzpicture}[->,>=stealth',shorten >=1pt,auto,node distance=3cm,
thick,main node/.style={circle,draw,font=\sffamily\Large\bfseries}]

\node[main node] (a) {$A$};
\node[main node] (b) [below left of=a] {$B$};
\node[main node] (c) [below right of=a] {$C$};

\path[every node/.style={font=\sffamily\small}]
(a) edge node  {$\frac{15}{40}$} (b)
edge [bend left] node {$\frac{12}{40}$} (c)
edge [loop above] node {$\frac{13}{40}$} (a)
(b) edge node  {$\frac{11}{40}$} (c)
edge [bend left] node {$\frac{5}{40}$} (a)
edge [loop left] node {$\frac{24}{40}$} (b)
(c) edge [bend left] node {$\frac{9}{40}$} (b)
edge node  {$\frac{8}{40}$} (a)
edge [loop right] node {$\frac{23}{40}$} (c)
;
\end{tikzpicture}
}
Consider the more extreme situation:\
$A\pref B \pref C : 15$ voters;\
$A \pref C \pref B : 4$ voters;\
$C\pref A \pref B: 6$ voters. This is in Figure \ref{fig:borda} on the left. 
According to the Borda count, in this case still $B$ wins. That is, although $C$ is preferred over $B$ by a large margin, $15:10$ the $6$ voters who bury $C$ under $A$ can override this and make $B$ winner. Our count still brings $C$ to be a winner, and only $9$ voters burying $C$ under $A$ could overturn this:  
$A \pref B \pref C : 15$ voters, $A \pref C\pref B : 1$ voter, $C\pref A\pref B:9$ voter, which is in Figure \ref{fig:borda} on the right.

The example shows that our technique is vulnerable to tactical voting, but -- as far as the effect of unpopular candidates go -- less so than the Borda count. To reduce the effect of irrelevant alternatives, some variants of Borda assign the points to candidates progressively, increasing the gap from lower to higher rank (cf., e.g., the Dowdall System~\cite{Reilly02dowdall}). We observe that convergence voting reduces the influence of irrelevant candidates in a more natural way.

%Our method still brings $C$ to be a winner, and only  $9$ voters could  overturn this: In the case of
%$A\pref B \pref C : 15$ voters;
%$A \pref C \pref B : 1$ voters;
%$C\pref A \pref B: 9$ voters,
%our count also makes $B$ the winner.

\ignore{\begin{tikzpicture}[->,>=stealth',shorten >=1pt,auto,node distance=3cm,
thick,main node/.style={circle,draw,font=\sffamily\Large\bfseries}]

\node[main node] (a) {$A$};
\node[main node] (b) [below left of=a] {$B$};
\node[main node] (c) [below right of=a] {$C$};

\path[every node/.style={font=\sffamily\small}]
(a) edge node  {$25$} (b)
edge [bend left] node {$19$} (c)
%edge [loop above] node {$24$} (a)
(b) edge node  {$15$} (c)
edge [bend left] node {$0$} (a)
%edge [loop left] node {$29$} (b)
(c) edge [bend left] node {$10$} (b)
edge node  {$6$} (a)
% edge [loop right] node {$22$} (c)
;
\end{tikzpicture}
}

The situation is even more striking if we increas the number of unpopular  candidates. For example, with
	\begin{itemize}
		\item  $E \pref D \pref A \pref B \pref C : 15 $
		\item   $E \pref D \pref A \pref C \pref B :  8 $
		\item     $C \pref E \pref D \pref A \pref B : 2 $
	\end{itemize}
	The Borda winner is again $B$. That is, although $C$ still wins over $B$ by $15:10$, this is overturned by only two voters who bury $C$ under the unpopular candidates $A$, $D$ and $E$. In other words the introduction of unpopular candidates multiply the voters capabilities to overturn the popular candidate $C$. In our method this effect is softened, we still need $4$ voters to overturn $C$:
	\begin{itemize}
		\item  $E \pref D \pref A \pref B \pref C : 15 $
		\item   $E \pref D \pref A \pref C \pref B :  6 $
		\item     $C \pref E \pref D \pref A \pref B : 4 $
	\end{itemize}

\ignore{\section{Rank Centrality}
	Rank Centrality was introduced in \cite{} as an estimator of the ranks of a set of players when
	only wins of pair-wise games are available. When for a given rank, the likelihood of winning pairwise games are
	given by the Bradley-Terry-Luce (BTL) model, then the Rank Centrality estimator of the rank has many good statistical properties. Our formula turns out to be exactly the same as Rank Centrality when a vote between a pair of candidates is considered to be a win of the preferred candidate in game, although we have arrived at the same formula through completely different considerations. In other words, if the BTL model is a good description of how the aptitude of the candidates for the post in question determines the probabilities of winning the voters' opinions, then Rank Centrality, and hence our way of ranking the candidates is a good estimator of how the candidates' true fitnesses to the post rank. This however is I think of secondary interest. The ranking method should give the best ranking based on the votes even if the votes are unreasonable and prefer bad candidates.
	
	In \cite{} the authors modify Rank Centrality to be applicable to Condorcet voting by making sure that the modified method choses the Condorcet winner when there is one. We argue here however, that although Rank Centrality might a different winner from the Condorcet winner, it is a good ranking.
}

\section{Conclusions}\label{sec:conclusions}

We have defined a new voting rule, called \emph{convergence voting}, and motivated by the PageRank algorithm. The idea is to use properties of Markov processes to rank candidates based on their pairwise comparisons, and then choose a winner. While our method does not necessarily produce the Condorcet winner, it can be seen as a simulation of a natural negotiating process in the community of voters. In this regard, we constructed a scheme to quantify the voters' community's support towards voting options based on pairwise preferences that gives the same Markov process as our voting scheme.  We have further shown that our voting scheme is Pareto efficient, does not allow dictators, but is not independent of irrelevant alternatives. Finally, we have compared our voting scheme with some of the well known other schemes such as Condorcet, Copeland and Borda and argued that  the output of convergence voting can be seen as a natural compromise between plurality and consensus voting rules.

While the new social choice function is mathematically similar to the Rank Centrality estimator, we want to emphasize that we are not interested in statistical estimation of some objectively existing ranking function for goods on the market or players in a tournament. We are solely interested in what kind of ranking is a good aggregation of the voters' preferences, to which the objective qualities of the voting options are entirely irrelevant.

%%%%%%%%%%%%%%%%%%%%%%%%%%%%%%%%%%%%%%%%%%%%%%%%%%%%%%%%%%%%%%%%%%%%%%%%%%%%%%%%%%%%%%%%%%%%%%%%%%%%%%%%%
% In the final version:
% - add acks of projects
% - thank David Naccache and Peter Roenne

%%%%%%%%%%%%%%%%%%%%%%%%%%%%%%%%%%%%%%%%%%%%%%%%%%%%%%%%%%%%%%%%%%%%%%%%%%%%%%%%%%%%%%%%%%%%%%%%%%%%%%%%%
%% bibliography: see CFP for number of permitted pages

%\bibliographystyle{plain}
\bibliography{wojtek,wojtek-own}  % put name of your .bib file here

\end{document}